\input{aipcheck}

\documentclass[final,numberedheadings]{aipproc}

\layoutstyle{6x9}

\def\lsim{\lower0.6ex\vbox{\hbox{$ \buildrel{\textstyle <}\over{\sim}\ $}}}
\def\gsim{\lower0.6ex\vbox{\hbox{$ \buildrel{\textstyle >}\over{\sim}\ $}}}

\def\beq{\begin{equation}}
\def\eeq{\end{equation}}

\def\Om{\Omega_{\rm M}}
\def\Ol{\Omega_{\rm \Lambda}}

\def\lcdm{$\Lambda\textrm{CDM}$~}

\def\himpc{{h$^{-1}$ Mpc}~}

\def\3he{$^3$He}
\def\4he{$^4$He}

\def\6li{$^6$Li}
\def\7li{$^7$Li}
\def\8Be{$^8$Be}
\def\9B{$^9$B}

\def\Msun{$M_{\odot}$~}

\def\Vmax{V_{\rm max}}

\def\dv2{\Delta_{\rm V/2}}
\def\Rv2{r_{\rm V/2}}

\def\sig8{\sigma_8}

\def\Mhost{M_{\rm host}}

\def\t0{t_0}

\def\ipl4{IPL4}

\begin{document}
\title{Halo Substructure and the Power Spectrum}

\author{Andrew R. Zentner}{
  address={Department of Physics, The Ohio State University, 174 W. 18th Ave., 
Columbus, OH 43210-1173}
}

\author{James S. Bullock}{
  address={Harvard-Smithsonian Center for Astrophysics, 60 Garden St., 
Cambridge, MA 02138}, altaddress={Hubble Fellow}
}

\begin{abstract}

In this proceeding, we present the results of a semi-analytic study of
CDM substructure as a  function of the  primordial power spectrum.  We
apply our method to several  ``tilted'' models in the \lcdm framework, 
with $n \simeq 1.1-0.8$, or $\sigma_8 \simeq 1.2 - 0.65$ when normalized to
COBE.  We also study a more extreme, warm dark matter-like 
spectrum that is sharply truncated below a scale of $\sim 0.3$ \himpc 
($\sim 10^{10}$ h$^{-1}$ M$_{\odot}$).
Contrary to some expectations, we show that the mass fraction 
of halo substructure is not a strong function of spectral slope, so 
it  likely  will be difficult  to constrain tilt using flux ratios of 
gravitationally lensed quasars. On the positive side,
all of our CDM-type models yield projected mass fractions
that  are in good   agreement  with  strong lensing
estimates: $f \approx 1.5\%$ at $M\sim10^8$\Msun.
The truncated model produces   a  significantly smaller fraction,  $f
\approx 0.3\%$, suggesting that  warm  dark matter-like spectra are 
disfavored and potentially may be distinguished from CDM spectra using lensing.
We also discuss  the issue of  dwarf satellite abundances, with 
emphasis on the cosmological dependence of the map between the observed
central velocity  dispersions of Milky  Way satellites and the maximum
circular  velocities of their host  halos.   In agreement
with earlier work, we find that standard \lcdm over-predicts
the estimated count of Milky Way satellites at fixed $\Vmax$ by an order
of magnitude, but   tilted  models  do better  because subhalos are less 
concentrated.   Interestingly, under the  assumption
that  dwarfs   have isotropic     velocity dispersion tensors, models    with
significantly tilted  primordial  power spectra ({\it e.g.}, $n  \lsim  0.85$,
$\sigma_8 \lsim 0.7$) \textit{may underpredict}  the number of  large
Milky Way satellites with $\Vmax \gsim 40$ km s$^{-1}$.

\end{abstract}
\maketitle


\section{Introduction}

The \lcdm model of a flat Universe dominated by cold, 
collisionless dark matter (CDM), and a cosmological constant ($\Lambda$) has 
emerged as the standard framework for the growth of cosmic structure.  With 
$\Om \approx 0.3$, $h \approx 0.7$, and a nearly scale-invariant primordial 
spectrum of adiabatic density perturbations ($P(k) \propto k^n$, $ n \approx 1$), 
\lcdm is remarkably successful at reproducing large scale observations.  In contrast, 
this paradigm faces several challenges on galactic and sub-galactic 
scales \cite{SAT,CD}.  In Zentner \& Bullock \cite{ZB} (ZB), 
we emphasized that inflation does not predict {\it exactly} scale-invariant 
({\it i.e.}, $n=1$) primordial spectra.  Many models of inflation predict 
``tilted'' spectra ($n \ne 1$), spectral index ``running'' 
($\textrm{d}n/\textrm{d}\ln k \ne 0$), or other deviations from 
scale-invariance that have dramatic consequences on small 
scales.  We showed that spectra with tilts of $n \sim 0.9$ and/or running 
and fixed by COBE on large scales can greatly reduce the predicted central 
densities of dark matter halos, alleviating the ``central density problem'' 
plaguing $\Lambda$CDM.  Further, the neighborhood of $\sig8 \sim 0.75$ implied 
by these tilts is provocatively close to many recent estimates  
of ``low'' $\sig8$ values \cite{SIG8}.  

In this proceeding, we report on results from follow-up work to ZB.  
We study the dependence of CDM halo substructure on the primordial 
power spectrum (PPS).  Our models of the PPS are the same 
as those in ZB.  We COBE normalize all spectra and we assume a cosmological 
model with $\Om = 1-\Ol = 0.3$, $\Omega_{\rm B}h^2 = 0.02$, and $h=0.72$.  
The important characteristics of each input spectrum are summarized in 
Table \ref{table:spectra}.  Numerical simulations cannot 
have both the resolution and the statistics needed to study substructure 
so we model substructure semi-analytically using host halo merger histories 
\cite{EPS} and a scheme for approximating subhalo orbits and tidal mass loss.  
Our model expands on previous work by Bullock {\it et al.} \cite{BKW} and Taylor 
and Babul \cite{TB}.  We calibrated our model against available data from N-body 
simulations; nevertheless, our results {\em must} be regarded as preliminary estimates 
to be verified by extensive N-body work.  We present results based on 100 merger 
tree realizations.  We give a detailed description of our model and further 
results in a forthcoming paper \cite{ZB2}. 
%
%
%
\begin{table}
\begin{tabular}{lcccc}
\hline
    \tablehead{1}{r}{b}{Model Description}
  & \tablehead{1}{r}{b}{Model Name}
  & \tablehead{1}{r}{b}{$n(k_{\rm COBE})$}
  & \tablehead{1}{r}{b}{$\textrm{d}n(k_{\rm COBE})/\textrm{d}\ln k$}
  & \tablehead{1}{r}{b}{$\sigma_{8}$}\\
\hline
Scale-invariant & $n=1$ & $\equiv 1$ & $\equiv 0$ & $\simeq 0.95$\\
Inverted Power Law & IPL4 & $\simeq 0.94$ & $\simeq -0.001$ & $\simeq 0.83$\\
Running-mass model I & RM I & $\simeq 0.84$ & $\simeq -0.004$  & $\simeq 0.65$\\
Running-mass model II & RM II & $\simeq 0.90$ & $\simeq -0.001$  & $\simeq 0.75$\\
Running-mass model III & RM III & $\simeq 1.1$ & $\simeq -0.001$ & $\simeq 1.21$\\
Broken scale-invariant & BSI & $=1.0$ & $=0$ & $\simeq 0.97$\\
\hline
\end{tabular}
\caption{Initial power spectra from the inflationary models discussed in ZB.}
\label{table:spectra}
\end{table}

\section{Substructure Mass Fractions}

Efforts have been made to use flux ratios in multiply-imaged 
quasars to detect substructure in galactic halos and to 
use these measurements to constrain cosmology.  In particular, Dalal and Kochanek 
\cite{DK} (DK) considered bounds on the PPS.  As such, it is important 
to understand the theoretical predictions for halo substructure as a 
function of the PPS and, more generally, substructure distributions 
and characteristics as a function of cosmology.  

\begin{figure}
  \includegraphics[height=.3\textheight]{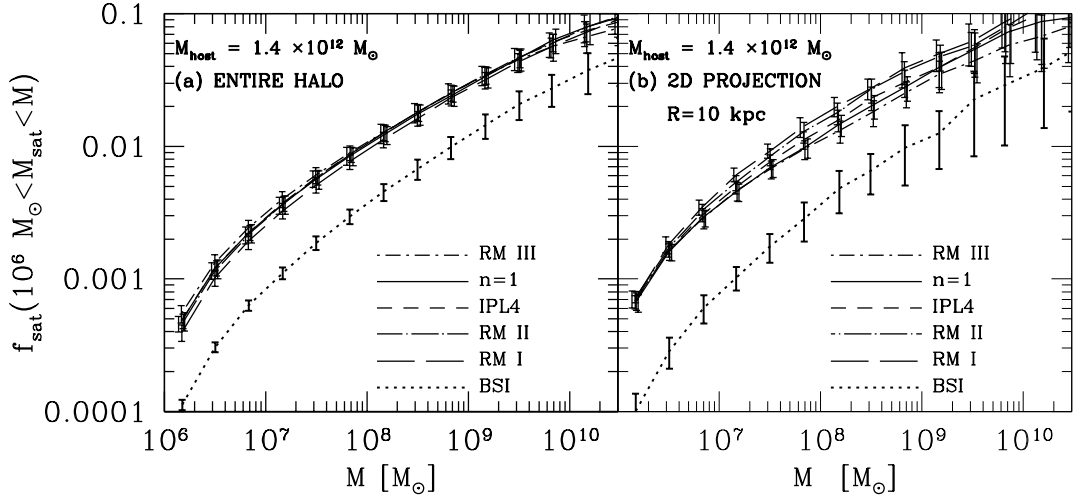}
  \caption{\label{fig:mfrac}The fraction of the host halo mass bound up in substructures 
of mass between $10^6 M_{\odot}$ and $M$ as a function of $M$. {\bf (a)} 
The mass fraction in substructure for the entire halo. {\bf (b)} 
The mass fraction in a 2D halo-centric cylindrical projection of 
radius $R=10$ kpc.  The lines represent the average mass fractions 
and the errorbars show the dispersion among the 100 realizations.  The models are 
labeled in each panel.}
\end{figure}

Our results on the substructure mass fraction and the PPS 
are summarized in Figure \ref{fig:mfrac}.  DK took a typical lens mass of 
$3 \times 10^{12}$ M$_{\odot}$ and the lenses in their sample have a median 
redshift of $z_{\ell} \approx 0.6$, so we present results for a 
$3 \times 10^{12}$ M$_{\odot}$ halo at $z=0.6$; however, our results do 
not change appreciably as a function of mass or redshift.  Lensing measurements 
are sensitive to the mass fraction in substructure projected onto the plane 
of the lens at a halo-centric distance of order the Einstein radius, 
$R_{\rm E} \sim 5$ kpc.  Consequently, we show in Fig. \ref{fig:mfrac} 
the mass fraction in substructure for the entire halo {\em and} the mass 
fraction in substructure in a 2D projection of radius $R = 10$ kpc.  

Notice that the substructure mass fraction is not a strong function 
of tilt and/or running.  In tilted models, host halos are less
concentrated and accrete their 
substructure later, and this compensates 
for the fact that the substructures 
are more fragile, and more easily destroyed by tides.  
It will be difficult to use substructure measurements 
to constrain these parameters.  Only the BSI model, with a sharp drop in power 
at $\sim 10^{10}$ M$_{\odot}$, shows deviation from the $n=1$ model that is 
significant compared to the scatter.  It may be possible to constrain 
models with such an abrupt break ({\it e.g.}, warm dark 
matter).  DK found the halo mass fraction bound up 
in substructures of mass $M_{\rm sat} \lsim 10^9$ M$_{\odot}$ to be 
$0.006 \lsim f_{\rm sat} \lsim 0.07$.  All of our models are consistent 
with this bound, but the truncated model is just at the edge of the allowed 
region.
%
%

\section{The Dwarf Satellites}

The ``dwarf satellite problem'', namely that \lcdm predicts roughly an
order of magnitude  more halos with  $\Vmax \lsim 40$ km s$^{-1}$ than
observed Milky Way (MW) satellites, is an often-discussed challenge to
\lcdm \cite{SAT}.  Stoehr  {\it et al.}  \cite{S02} (S02)  and Hayashi
{\it et al.} \cite{H02} (HO2)  proposed that substructure halos may be
significantly less   concentrated than comparable   field halos due to
tidal  effects.   This implies   that   the values  of  $\Vmax$   that
correspond    to   the    observed  central    velocity   dispersions,
$\sigma_{\star}$,  of the MW satellites   are  larger than the  values
inferred by  other authors.  One  must  be cautious.  Mass redistribution 
in subhalos is quite sensitive to the mass resolution of the simulation 
(S02) and the velocity function (VF) of satellite halos is sensitive to the 
initial concentrations and accretion times of substructure (H02).  Our  
semi-analytic model represents one extreme; we do not allow  for 
redistribution of mass within a subhalo's  tidal  limit.  Using  our model, 
we can also quantify the cosmology dependence  of the mapping between 
$\sigma_{\star}$ and $\Vmax$.  We have assumed that CDM halos can be well 
described by NFW \cite{NFW} profiles with a  particular $\Vmax$ and $R_{\rm  max}$ (the
radius at which $\Vmax$ is  attained) and calculated all  combinations
of $\Vmax$ and   $R_{\rm max}$ that lead  to  the observed values   of
$\sigma_{\star}$ for each of the MW satellites.   We have assumed that
the stars have   isotropic dispersion  tensors  and  that  the stellar
distributions are given by  King profiles \cite{KING} with  parameters
given by Mateo \cite{MATEO}.
%
%
\begin{figure}
  \includegraphics[height=.3\textheight]{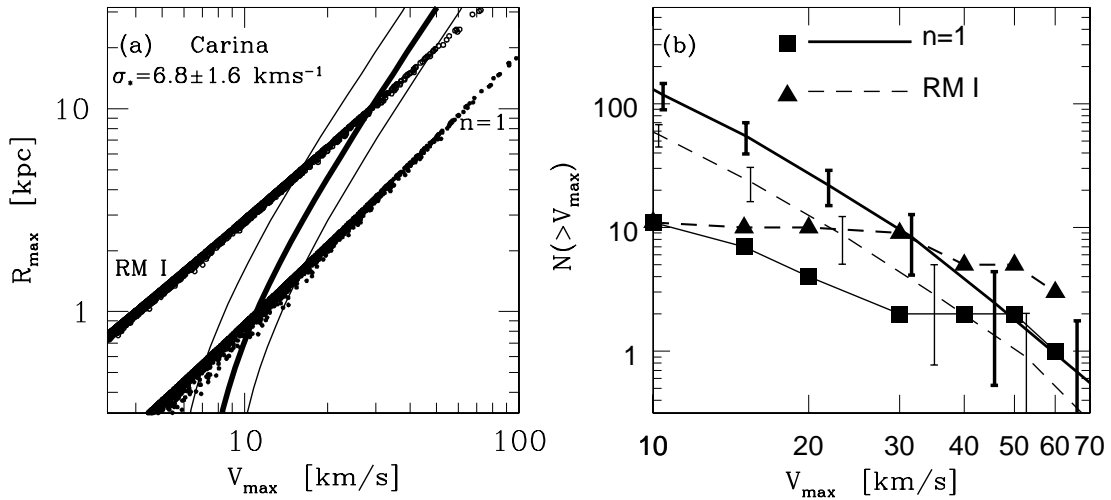}
  \caption{\label{fig:vm-rm}{\bf  (a)}  The   lower   group  of points
  represent a  scatter   plot of  $\Vmax$  vs.  $R_{\rm max}$  for  10
  realizations of the $n=1$ model.   The upper points correspond to RM
  I.   The lines  show  the region that  yields  an observed value  of
  $\sigma_{\star} = 6.8 \pm 1.6$  km  s$^{-1}$ for Carina.  The  thick
  solid line  corresponds to  the   central value of  $\sigma_{\star}$
  while the  thin  solid  lines correspond  to   the $1\sigma$ errors.
  Consistency   demands  that  Carina  resides in    a   halo that has
  structural properties that lie in the  region of overlap between the
  thin solid lines and the scattered points  for each cosmology.  {\bf
  (b)} The predicted VFs  (lines) and scatter for the  $n=1$ and RM  I
  models   along  with the ``observed'' VF    (shapes)  for each model
  inferred from the observed values of $\sigma_{\star}$.  Squares
   represent satellite velocities that would be implied if their
   halo profiles reflect halos in the $n=1$ model, triangles correspond
   to the RM I expectations.}
\end{figure}

Results  for Carina are shown in  Figure \ref{fig:vm-rm}, along with a
scatter plot of $\Vmax$ vs. $R_{\rm max}$ for the surviving satellites
in 10 realizations of a $\Mhost = 1.4 \times 10^{12}$ M$_{\odot}$ host
halo  at $z=0$ for the   $n=1$ and RM  I  models. 
This plot shows how
allowing  for  less concentrated halos  helps  to  alleviate the dwarf
satellite problem.  Less  concentrated  halos  (larger $R_{\rm  max}$)
require   a  larger $\Vmax$  because the outer stellar radius of
Carina is much smaller than the radius at which the halo's rotation
curve peaks.  Larger   halos are intrinsically   scarcer
objects, helping to explain the paucity of dwarf satellites.
Feedback  mechanisms ({\it  e.g.}, \cite{BKW})  then explain
the dearth of smaller halos.  Figure \ref{fig:vm-rm} also demonstrates
that  the mapping between   $\sigma_{\star}$ and $\Vmax$  is dependent
upon cosmology, in particular the PPS, and  so the same
observational data  imply  a  {\em  cosmology dependent}  ``observed'' 
VF.  Our estimates for dwarf velocities in 
the $n=1$ case compare well to
the estimates made by Klypin and collaborators \cite{SAT}.

In the right panel of Fig. \ref{fig:vm-rm} 
we present the predicted VFs along with separate ``observed'' velocity 
functions for the $n=1$ and RM I models.  The RM I VF is a factor of 
$\sim 2$ lower than the $n=1$ VF mainly because typical halos are 
less concentrated in this model, so that 
$\Vmax$ is lower at a given mass.  Also notice that the ``observed'' VF 
is shifted significantly higher at high $\Vmax$.  This suggests that 
significantly tilted power spectra $n \lsim 0.85$ may {\em underpredict} 
the number of MW satellites at high $\Vmax$.  Moreover, subhalos are 
likely less centrally concentrated than field halos 
(as suggested by S02 and H02), and this serves to make the 
underprediction {\em more} pronounced.  However, we 
recommend circumspection.  Our results concerning VFs are 
sensitive to several assumptions such as the isotropy of the 
dispersion tensor.
%
%
%

\bibliographystyle{aipproc}

\end{document}